# Decentralized Cooperative Lane Changing at Freeway Weaving Areas Using Multi-Agent Deep Reinforcement Learning

Yi Hou, Peter Graf

*Abstract*— Frequent lane changes during congestion at freeway bottlenecks such as merge and weaving areas further reduce roadway capacity. The emergence of deep reinforcement learning (RL) and connected and automated vehicle technology provides a possible solution to improve mobility and energy efficiency at freeway bottlenecks through cooperative lane changing. Deep RL is a collection of machine-learning methods that enables an agent to improve its performance by learning from the environment. In this study, a decentralized cooperative lane-changing controller was developed using proximal policy optimization by adopting a multi-agent deep RL paradigm. In the decentralized control strategy, policy learning and action reward are evaluated locally, with each agent (vehicle) getting access to global state information. Multi-agent deep RL requires lower computational resources and is more scalable than single-agent deep RL, making it a powerful tool for time-sensitive applications such as cooperative lane changing. The results of this study show that cooperative lane changing enabled by multi-agent deep RL yields superior performance to human drivers in term of traffic throughput, vehicle speed, number of stops per vehicle, vehicle fuel efficiency, and emissions. The trained RL policy is transferable and can be generalized to uncongested, moderately congested, and extremely congested traffic conditions.

## I. Introduction

LANE changing is one of the most important and frequent driving maneuvers, involving both longitudinal and lateral movement with interactions with other vehicles. Successful lane changes require drivers to account for several safety-related factors, including speed and position of the ego vehicle and vehicles surrounding it, road geometries, and traffic volume. Inaccurate assessment of these factors leads to car accidents, congestion, waste of energy, and poor air quality. Approximately 5% of all crashes and as high as 7% of all crash fatalities are related to improper lane-changing maneuvers [1]–[3]. Additional technologies such as blind spot warning, closing vehicle warning, and lane-change warning have been deployed on automobiles to prevent bad lane changes in the last few decades. The emergence of connected and automated vehicles offers a great opportunity to improve lane changing. With sensing technology and cutting-edge vehicle-to-everything (V2X) communication, vehicles can make lane-change maneuvers through cooperation. Cooperative lane changing has been demonstrated to be effective in mitigating congestion and reducing energy consumption and delay with smoothed traffic flow [4]–[5].

This paper develops a decentralized controller for cooperative driving at freeway weaving areas using multi-agent deep reinforcement learning (RL), as frequent lane changes at freeway bottlenecks such as merging and weaving areas further reduce roadway capacity [6]. Recently, RL has emerged as a group of methods to learn optimal control policies in complex, nonlinear, and stochastic environments. RL directly interacts with the environment based on the rewards it receives to iteratively improve upon control policies. RL has been successfully applied to autonomous driving, and these studies can be summarized into two categories. In the first, researchers only focus on high-level decision-making in vehicle control, such as lane-change decisions and acceleration [7]–[10]. In the second category, the studies mainly focus on end-to-end vehicle control, including route planning, steering control, and acceleration [11]–[13].

Recently, there has been growing interest in applying RL to cooperative lane changing. Wang et al. [14] proposed a cooperative lane-changing model that only outputs discrete lane-changing decisions without controlling the longitudinal movement using multi-agent RL. The simulation results showed that cooperative lane changing led to smoother traffic flow. Dong et al. [15] developed a centralized cooperative lane-changing controller using a deep Q network combined with a graph convolutional neural network. They constructed a graph between subject vehicle and neighbor vehicles to obtain global state information as inputs to the deep Q network. The RL framework only outputs discrete lane-changing decisions as actions. Ren et al. [16] proposed a cooperative lane-changing framework for work zone merge control using RL. The framework only controls longitudinal movement of merging vehicles to find safe lane-changing gaps. The results showed that cooperative merge control outperformed human drivers in

Manuscript was submitted on October 5th, 2021 for review. This work was supported in part by the U.S. Department of Energy Vehicle Technologies Office (VTO) under the Energy Efficient Mobility Systems (EEMS) Program.

The authors are with the National Renewable Energy Laboratory, 15031 Denver W Pkwy, Golden, CO 80401 USA. (e-mail: Yi.Hou@nrel.gov; Peter.Graf@nrel.gov)

terms of both mobility and safety. These studies developed RL models that control either only lane-change decision-making or only vehicle longitudinal movement for gap searching. In this paper, we propose using multi-agent deep RL to control both lane-changing decisions and longitudinal movement. Including longitudinal movement in lane-change control has the potential to further improve roadway capacity, as longitudinal movement, such as slowing down and speeding up, enables more coordination between vehicles to proactively create gaps for lane changing. We assume that all vehicles are fully automated and connected and adopt a decentralized control strategy where policy learning and action reward are evaluated locally, with each agent (vehicle) getting access to global state information. Decentralized control requires less computational resources and enables faster RL training than a centralized controller. Moreover, decentralized control is more likely to be deployed in the future because any form of liability would be limited to the single agent (vehicle).

The remainder of the paper is organized as follows: Section II provides a detailed description of the multi-agent deep RL method we propose for cooperative lane changing. Section III describes the experiment setup for both simulation and the multi-agent deep RL training process. Section IV presents the RL results and compares them with human drivers, and Section V summarizes the study and discusses conclusions and future research.

## II. METHODOLOGY

### A. Reinforcement Learning

Cooperative lane changing is modeled as a fully observable Markov decision process in this study. In Markov decision process formulation, given the current state ($s_t \in \mathcal{S}$) and a control policy ($\pi: S \rightarrow A$), an agent performs an action from a set of possible actions ($a_t \in \mathcal{A}$) at each step in a control horizon ($t \in \mathcal{T}$) and receives an immediate reward, $r_t$, from the environment based on the action. The objective of RL is to obtain the optimum policy $\pi^*$ from its learning experiences to maximize the expected cumulative discounted reward from the environment (i.e., $\mathbb{E}_\pi(\sum_{t \in \mathcal{T}} \gamma^t r_t)$), where $\gamma \in (0,1)$ is the discount factor.

RL methods can be divided into three main categories depending on the form of the control policy $\pi$. They are value-based, policy-based, and actor-critic methods. For value-based RL, value functions are used to represent the expected future reward given the current state and action. For example, Q-function is a type of value function. The Q-function of taking action $a$ while in state $s$ following policy $\pi$ is defined as [17]:

$$Q_\pi(s,a) = \mathbb{E}_\pi \left[ \sum_{\tau=t}^{|\mathcal{T}|} \gamma^{\tau-t} r_\tau \,|\, s_t = s, a_t = a \right]. \quad (1)$$

A greedy policy is obtained with $Q_\pi(s,a)$ learned via approaches such as Monte Carlo learning, state–action–reward–state–action (SARSA) learning [18], and Q-learning. Value-based RL is implicitly defined through the following maximization problem (2):

$$a_t = \pi(s_t) = \underset{a \in \mathcal{A}}{\mathrm{argmax}}\, Q(s_t, a). \quad (2)$$

Policy-based RL directly optimizes the parameterized policy (i.e., $\pi_\theta(s_t)$, where $\boldsymbol{\theta}$ is the parameter vector) instead of the value function. Consequently, the control performance depending on $\pi_\theta$ becomes a function of $\boldsymbol{\theta}$. The optimal parameter $\boldsymbol{\theta}^*$ resulting in the optimal policy is obtained by applying stochastic gradient descent at each learning iteration until convergence, as shown in (3). The gradient term $\widehat{\nabla}_\theta J(\boldsymbol{\theta})$ is estimated from experience collected in the current learning iteration.

$$\boldsymbol{\theta}^{t+1} = \boldsymbol{\theta}^t + \alpha \widehat{\nabla}_\theta J(\boldsymbol{\theta}) = \boldsymbol{\theta}^t + \alpha \widehat{\nabla} \mathbb{E}_{\pi_\theta} \left[ \sum_{t \in \mathcal{T}} \gamma^t r_t \right] \quad (3)$$

Actor-critic RL is a hybrid method learning both policy (actor) and value functions (critic). The critic in actor-critic approaches generates lower variance when estimating $\widehat{\nabla}_\theta J(\boldsymbol{\theta})$ compared with pure policy-based algorithms [19].

### B. Proximal Policy Optimization

In this study, we choose the actor-critic algorithm, proximal policy optimization (PPO) [20], due to its popularity and successful implementations demonstrated in the past decade. For policy-based approaches, in practice, large policy updates (coinciding with gradients in (3) of large norm) might occur and cause a destructive effect during learning. To mitigate the risk of detrimental policy shifts, a trust region policy optimization (TRPO) algorithm [21], which constrains the Kullback–Leibler divergence between the policies before and after the update, was developed. Though TRPO has proven to be effective, its implementation is complicated. As a result, the PPO algorithm was proposed, which inherits the benefits provided by TRPO but has much simpler implementation. Specifically, the objective function used in PPO is given by:

$$\mathcal{L}(\boldsymbol{\theta}) = \mathbb{E}_t[\min(\kappa_t(\boldsymbol{\theta})A_t, clip(\kappa_t(\boldsymbol{\theta}), 1-\epsilon, 1+\epsilon)A_t)], \quad (4)$$

where $\kappa_t(\boldsymbol{\theta}) = \frac{\pi_\theta(a_t|s_t)}{\pi_{\theta_{old}}(a_t|s_t)}$, $A_t$ is an estimation of the advantage function, and $\epsilon$ is a hyperparameter. The rationale behind this objective function is to penalize updates of $\pi_\theta$ that move $\kappa_t(\boldsymbol{\theta})$ away from 1, thus considering the Kullback–Leibler divergence of policy updates without explicitly formulating the constraint in the optimization problem.

## III. EXPERIMENTS

### A. Experiment Setup

To experiment the proposed multi-agent deep RL framework, a typical three-lane freeway weaving area with an on-ramp and an off-ramp was constructed in the simulation environment using Simulation of Urban MObility (SUMO) [22], as shown in Fig. 1. The freeway segment is 500 meters long in total with 200 meters upstream of the on-ramp and 100 meters downstream of the off-ramp. The speed limit on the freeway is

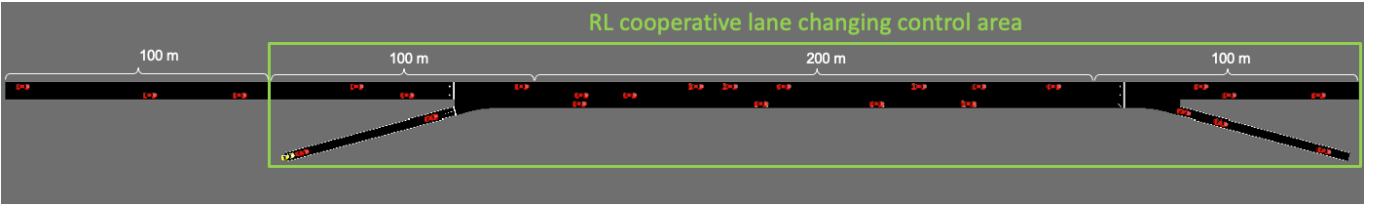

**Fig. 1.** An on-ramp merging area on a two-lane freeway in a SUMO simulator.

65 miles per hour (mph) and 40 mph on ramps. FLOW [23], an open-source RL framework that integrates SUMO and RLlib [24], was used for RL training and evaluation. Because the main purpose of cooperative lane changing is to alleviate traffic congestion and improve roadway capacity at freeway merging and weaving areas, we needed to create enough opportunities for vehicles to change lanes in congested traffic conditions during RL training. Moderate congestion was observed when the traffic inflow of both the freeway and on-ramp was set at 1,200 vehicles per hour per lane (vphpl), with half of the vehicles from freeway inflow exiting through the off-ramp and all vehicles from on-ramp inflow entering the freeway. Thus, the RL policy is trained with traffic inflow of 1,200 vphpl. Vehicles do not initiate cooperative lane changing until 100 meters upstream of the on-ramp and do not end cooperative lane changing until 100 meters downstream of the off-ramp. The cooperative lane-changing control area is highlighted in Fig. 1. For each simulation episode, we ran 1,000 time steps with each time step as 0.2 seconds. Because the RL cooperative lane-changing control policy developed in the paper is to be applied to autonomous vehicles that can drive safely on the road, the main focus of this study is to train vehicles for cooperative lane changing instead of training them how to drive safely. Thus, we used the default collision prevention mechanism in SUMO to prevent vehicle collisions during simulation.

### B. Action and State Space

The actions of each vehicle are acceleration and lane-changing decision. Acceleration is a real value with negative values representing deceleration. The maximum acceleration is 4 $m/s^2$ and the maximum deceleration is 8 $m/s^2$. The lane-changing decision is a discrete variable including decisions of staying in the current lane, changing to the left lane, and changing to the right lane.

The state space consists of the state of the subject vehicle and its neighbor vehicles. The state of the subject vehicle includes speed, position coordinates, current lane, and destination (exiting through the off-ramp or staying on the freeway). A total of six neighbor vehicles are observed by the subject vehicle in Fig. 2. They are the leading and following vehicles on the same lane as the subject vehicle and the adjacent vehicles in front of and behind the subject vehicle in both the left and right lanes. The state of neighboring vehicles includes longitudinal distances to subject vehicle, speeds, blinker statuses, and destinations (exiting through the off-ramp or staying on the freeway). It is assumed that the vehicle sensor detection range is 200 meters. If there is no vehicle within the detection range, the neighbor vehicle state space values still need to be filled. If there is no leading vehicle or adjacent vehicles in front of the subject vehicle in both the left and right lanes within the detection range, we fill speed with speed limit value, distance with 200 meters, blinker status with off, and destination with staying on freeway. If there is no following vehicle or adjacent vehicles behind the subject vehicle on both the left and right lanes within the detection range, we fill speed with 0, distance with 200 meters, blinker status with off, and destination with staying on freeway. If there is no lane on the left or right side of the subject vehicle, the state spaces of all neighbor vehicles on these lanes are filled with 0. State space values are normalized between 0 and 1 to accelerate the RL training process.

### C. Reward Function

The reward function is a key component to the success of RL training. The optimal policy is defined by the reward function together with the action space, because the agent chooses the action that maximizes the expected discounted reward at every time step. The reward function in this study was designed by considering mobility, safety, and comfortability. The details of the three components are discussed below.

- **Mobility.** The cooperative lane-changing controller should be designed to enable successful lane changes without slowing down traffic. Thus, the reward function should reward high vehicle speed and successful lane changes.
- **Safety.** Safety is of paramount importance for any vehicle operations. In order to minimize the chance of vehicle collision, the reward function should be designed to penalize small headways, improper lane change intentions, and emergency brakes. Improper lane change is defined as any lane change that can cause a collision with other vehicles in the target lane. Emergency brakes are brakes that cause deceleration greater than 9 $m/s^2$.

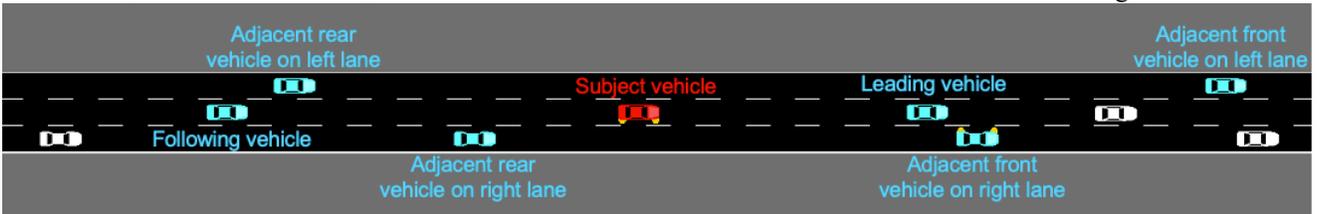

**Fig. 2.** Subject vehicles and neighbor vehicles.

- **Comfortability.** The control algorithm needs to prevent unnecessary lane changes to achieve comfortability for passengers. Thus, the reward function penalizes any unnecessary excessive lane-changing maneuver.

More specifically, the reward function is defined as:

$$R(s_t, a_t) = \sum_{i \in V_t} W_v \cdot v_i + W_l \cdot l_i + W_c \cdot c_i + W_s \cdot s_i + W_b \cdot b_i + W_h \cdot h_i, \quad (5)$$

where

$s_t$ is the state space observed by all vehicles engaging in cooperative lane changing at time $t$.
$a_t$ is the action performed by all vehicles engaging in cooperative lane changing at time $t$.
$V_t$ is the set of vehicles engaging in cooperative lane changing at time $t$.
$v_i$ is the speed of vehicle $i$.
$l_i$ is the reward (penalty) for vehicle $i$ staying (not staying) on the desired lanes.
$c_i$ is the penalty for any lane-change maneuver performed by vehicle $i$.
$s_i$ is the penalty for improper lane-change intention of vehicle $i$.
$b_i$ is the penalty for emergency brake performed by vehicle $i$.
$h_i$ is the penalty for small time headway of vehicle $i$.
$W_v$ is the weight coefficient for the speed reward.
$W_l$ is the weight coefficient for the reward (penalty) of staying (not staying) in desired lanes.
$W_c$ is the weight coefficient for the lane-changing penalty.
$W_s$ is the weight coefficient for the penalty of improper lane-changing intention.
$W_b$ is the weight coefficient for the emergency brake penalty.
$W_h$ is the weight coefficient for the small time headway penalty.

Making earlier decisions to change lanes leads to a larger chance to make successful lane changes with minimum negative impact on traffic, because vehicles have more time make speed adjustment to create gaps for lane changing. Last-minute lane changes approaching the end of the weaving area usually cause congestion by either forcing vehicles in the target lane to slow down or blocking the current lane while waiting for an opportunity to change lanes. Therefore, the reward $l_i$ is designed to encourage early lane changes. The reward for staying in desired lanes should decrease when approaching the end of weaving area, whereas the penalty for not staying on the desired lanes should increase. We specify $l_i$ as

$$l_i = \begin{cases} 1 - \dfrac{d_i}{d_{max}}, & \text{if vehicle } i \text{ is on target lane} \\ -\dfrac{d_i}{d_{max}}, & \text{otherwise,} \end{cases} \quad (6)$$

where

$d_i$ is the distance between vehicle $i$ and on-ramp entrance.
$d_{max}$ is the distance between on-ramp entrance and off-ramp exit.

Smaller time headway is more likely to cause collision, but the relationship between vehicle time headway and likelihood of collision is not linear. When time headway is above some threshold, increasing time headway does not further lower risk of collision. Therefore, the penalty for small time headway $h_i$ is defined as

$$h_i = \min\left(\frac{t_i - t_{min}}{t_{min}}, 0\right), \quad (7)$$

where

$t_i$ is the time headway of vehicle $i$.
$t_{min}$ is the minimum time headway and is set as 1 second, which is the minimum headway of the default SUMO lane-following model.

Emergency brake penalty $b_i$, lane change penalty $c_i$, and improper lane-change intention penalty $s_i$ are defined as binary variables as below:

$$b_i = \begin{cases} -1, & \text{if vehicle } i \text{ performs emergency brake} \\ 0, & \text{otherwise} \end{cases} \quad (8)$$

$$c_i = \begin{cases} -1, & \text{if vehicle } i \text{ changes lane} \\ 0, & \text{otherwise} \end{cases} \quad (9)$$

$$s_i = \begin{cases} -1, & \text{if vehicle } i \text{ intents to perform} \\ & \text{improper lane change} \\ 0, & \text{otherwise.} \end{cases} \quad (10)$$

A trial-and-error process was conducted to determine the optimum weight coefficients $W_v$, $W_l$, $W_c$, $W_s$, $W_b$, and $W_h$. It was found that $W_v = 0.1$, $W_l = 1$, $W_c = 1$, $W_s = 5$, $W_b = 1$, and $W_h = 1$ yields the best performance.

## IV. EXPERIMENT RESULTS AND ANALYSIS

We ran experiments on 32 central processing units with a multi-agent RL paradigm where a shared RL control policy is trained with multiple agents interacting with each other in a common environment. A few hyperparameters of PPO including sample size of each training iteration, learning rate, and neural network structure were tuned for optimum model performance. PPO produced the best results with a sample size of 16,000 time steps per training iteration and a learning rate of 5e−5. A neural network with one hidden layer of 128 neurons and a tanh activation function was used for training. Fig. 3 presents the reward curve of PPO over the training cycle. The reward curve plateaus after about 3 million training steps (about 60 wall clock hours).

The trained PPO cooperative lane-change control policy was first evaluated with the same traffic inflow setting, 1,200 vphpl, as during RL training, which generates moderate congestion. It was then further tested in extremely congested traffic conditions where traffic inflow was set as constant 1,500 vphpl and uncongested traffic conditions where traffic inflow was set as constant 900 vphpl. The results were analyzed from perspectives of mobility, energy, and emissions. The simulation results include the following metrics:

- Traffic throughput (vph)
- Average vehicle speed (m/s)
- Average number of stops per vehicle
- Average vehicle fuel efficiency (mpg)
- Average $CO_2$ (g/mi) emissions per vehicle
- Average $NO_x$ (mg/mi) emissions per vehicle.

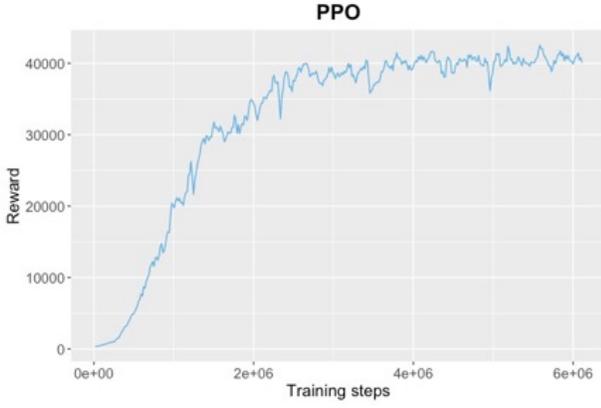

**Fig. 3.** Reward curves of PPO.

The default energy and emissions model in SUMO, HBEFA3/PC_G_EU4 [25], is used for energy and emissions estimation in this study. We ran 30 simulation episodes for the trained multi-agent deep RL policy and human drivers to achieve statistical significance for the performance results.

The comparison of results between the multi-agent deep RL cooperative lane-changing controller and human drivers is displayed in Fig. 4. The error bar is used to show the standard deviation of the performance metrics and the circle is used to show the mean value. As shown in Fig. 4, the multi-agent deep RL cooperative lane-changing controller yielded higher mean values in traffic throughput and fuel efficiency than the human drivers, and lower mean values in travel time, number of stops, and emissions in scenarios of 1,200 vphpl and 1,500 vphpl, indicating that multi-agent, deep-RL-enabled cooperative lane changing outperformed human drivers in moderately congested and extremely congested scenarios. For uncongested traffic condition when traffic inflow is 900 vphpl, the performance of cooperative driving is comparable with human drivers with slightly reduced number of stops and emissions and slightly increased fuel efficiency, but slightly increased travel time and slightly reduced throughput. As there are few interactions between vehicles during lane changing in uncongested traffic conditions, the advantage of cooperation is diminished. The reason for longer travel time resulting from cooperative driving in the uncongested scenario might be that vehicles learned to drive more conservatively in congested scenarios and were therefore not as aggressive as human drivers when transferring the policy to uncongested traffic conditions. Moreover, the multi-agent deep RL cooperative lane-changing controller tends to be more stable than human drivers, as cooperative lane changing in general has less variance in performance metrics compared with human drivers.

The percentage differences between the results of the multi-agent deep RL cooperative lane-changing controller and human drivers are shown in Table I. For moderately congested and extremely congested scenarios, cooperative lane changing, on average, respectively increased traffic throughput by 6.2% and 14.5%; reduced vehicle travel time by 29.0% and 42.0%; reduced number of stops per vehicle by 97.0% and 91.5%;

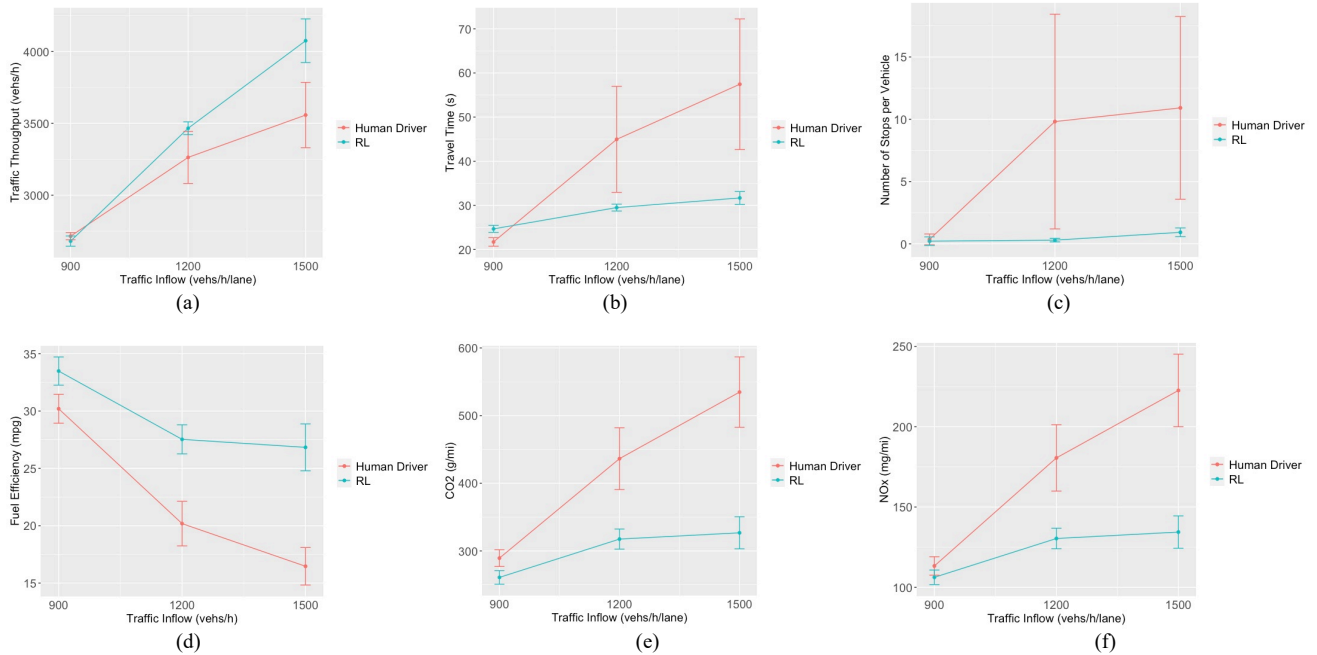

**Fig. 4.** Comparison of multi-agent deep RL algorithms and human driver performance with different traffic inflow for (a) traffic throughput, (b) vehicle speed, (c) number of stops per vehicle, (d) fuel efficiency, (e) $CO_2$ emissions, and (f) $NO_x$ emissions.

TABLE I
PERFORMANCE METRIC CHANGES WHEN COMPARED WITH HUMAN DRIVERS
IN DIFFERENT TRAFFIC CONDITIONS

| | No Congestion (Inflow: 900 vphpl) | Moderate Congestion (Inflow: 1,200 vphpl) | Extreme Congestion (Inflow: 1,500 vphpl) |
|---|---|---|---|
| Traffic throughput (vph) | −1.3% | +6.2% | +14.5% |
| Average travel time (sec) | +13.7% | −29.0% | −42.0% |
| Average number of stops per vehicle | −39.3% | −97.0% | −91.5% |
| Average vehicle fuel efficiency (mpg) | +10.9% | +36.4% | +63.0% |
| Average $CO_2$ emissions per vehicle (g/mi) | −9.9% | −27.2% | −38.9% |
| Average $NO_x$ emissions per vehicle (mg/mi) | −6.2% | −27.8% | −39.6% |

increased fuel efficiency by 36.4%, and 63.0%; reduced $CO_2$ emissions by 27.2% and 38.9%; and reduced $NO_x$ emissions by 27.8% and 39.6%. For the uncongested scenario, cooperative lane changing, on average, had 1.3% less traffic throughput, 13.7% higher travel time, 39.3% fewer number of stops per vehicle, 10.9% higher fuel efficiency, 9.9% less $CO_2$ emissions, and 6.2% less $NO_x$ emissions than human drivers.

Traffic density was further analyzed and compared between multi-agent deep RL and human drivers in uncongested, moderately congested, and extremely congested traffic conditions. Traffic density was calculated by aggregating the number of vehicles to 10-meter segments at every simulation time step. Fig. 5 displays the traffic density map of the cooperative lane-changing control area on the freeway segments for both multi-agent deep RL controllers and human drivers. The x-axis is the freeway longitudinal driving distance, and the y-axis is the elapsed time. Lighter colors represent higher traffic density, whereas darker colors represent lower traffic density. When vehicles are evenly distributed on the freeway segment and no queue was built up, there are similar colors across the density map with little color changes, implying smooth traffic flow. Free-flow traffic conditions result in darker colors across the map, whereas heavy congestion results in light-colored patches moving upstream as time elapses because of traffic shock wave propagation.

Comparing Fig. 5(a) and 5(d) shows that when traffic inflow is 900 vphpl, the multi-agent deep RL cooperative lane-changing controller has slightly higher traffic density across the freeway segments but with fewer queues built up. Fig. 5(a) shows that there is a short queue developed at around 300 meters at roughly 120 seconds, whereas Fig. 5(d) shows that four short queues were developed at around 300 meters at four different times. Comparing Fig. 5(b) and 5(e) demonstrates that when traffic inflow is 1,200 vphpl, cooperative lane changing has much smoother traffic than human drivers. Fig. 5(e) indicates vehicles evenly spread on the freeway without congestion, whereas Fig. 5(b) shows a long queue has been built up at around 300 m at 75 seconds, and it propagates upstream until 150 seconds. Comparison between Fig. 5(c) and 5(f) also shows that cooperative lane changing has much smoother traffic than human drivers in extremely congested traffic conditions when traffic inflow is 1,500 vphpl. Fig. 5(f) shows very smooth color change, indicating low speed variance and smooth traffic, whereas Fig. 5(c) shows traffic congestion shock wave propagation occurring several times during the simulation. In general, the traffic density map comparison demonstrates that the multi-agent deep RL cooperative lane-changing controller is very effective in smoothing traffic and mitigating traffic congestion in different traffic conditions.

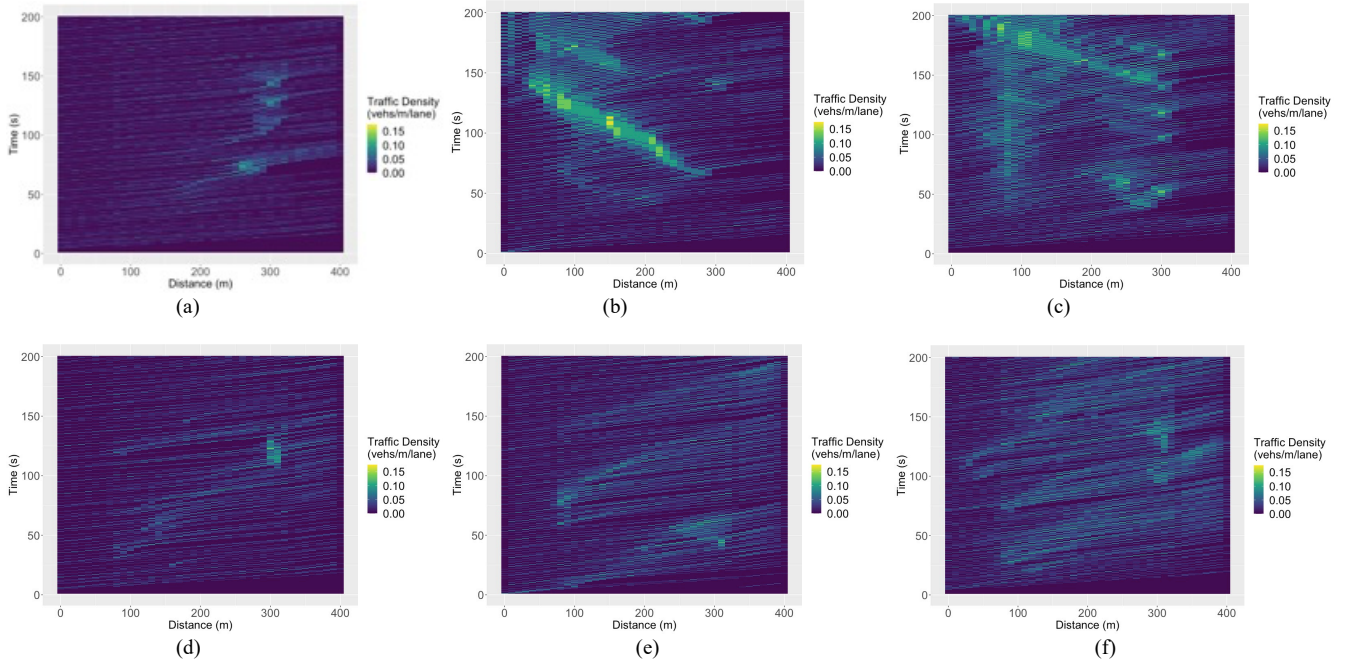

**Fig. 5.** Traffic density map for (a) human drivers with inflow of 900 vphpl (b) human drivers with inflow of 1200 vphpl (c) human drivers with inflow of 1500 vphpl (d) cooperative lane changing with inflow of 900 vphpl (e) cooperative lane changing with inflow of 1200 vphpl and (f) cooperative lane changing with inflow of 1500 vphpl

A more intuitive illustration of the shock waves caused by congestion is to plot the vehicle trajectories. The vehicle trajectories of the cooperative lane-changing control area on freeway segments for both multi-agent deep RL controllers and human drivers are plotted in Fig. 6. Each curve represents a trajectory, with green indicating higher speed and red indicating lower speed. The comparisons between Fig. 6(a) and 6(d), Fig. 6(b) and 6(e), and Fig. 6(c) and 6(f) demonstrate that the multi-agent deep RL cooperative lane-changing controller is effective in eliminating stop-and-go traffic and its associated shock waves in uncongested, moderately congested, and extremely congested conditions.

## V. Conclusions and Future Work

In this study, multi-agent deep RL was used to train control policies for cooperative lane changing at a freeway weaving area in the SUMO simulator using Flow and RLlib. The results show that the trained multi-agent deep RL cooperative lane-changing controller outperforms human drivers in terms of traffic throughput, travel time, number of stops per vehicle, vehicle fuel efficiency, and $CO_2$ and $NO_x$ emissions in uncongested, moderately congested, and extremely congested traffic conditions. This paper demonstrates that cooperative lane changing enabled by multi-agent deep RL is effective in smoothing traffic, reducing speed variance, and eliminating traffic shock waves.

Multi-agent deep RL is an ideal tool for solving the cooperative lane-changing problem because of its ability to learn policies from stochastic simulation episodes representing an uncertain, time-dependent environment within which control decisions are made. Cooperative lane changing involves the need to make optimal control decisions for each individual vehicle across time in a time-evolving environment with an uncertain trajectory. Instead of formulating and solving an explicit optimization model as classical optimization problems, RL finds good solutions via repeatedly evaluating RL-based decisions via forward simulation and then using reward feedback from the simulation to incrementally improve the policy. These features allow RL to be deployed in time-sensitive applications, such as cooperative lane changing, without reoptimizing from scratch every time traffic conditions change by training a policy in the context of stochastic scenarios that reflect the expected, dynamic behavior of the real-world system.

This study only evaluated the effectiveness of deep RL in cooperative lane-changing control in an environment where all vehicles are autonomous vehicles. In future research work, cooperative lane changing in mixed traffic with both autonomous vehicles and human drivers will be explored. Future research efforts will also focus on designing reward functions for more explicit cooperation. Transferring an RL policy trained on a simpler simulator to a more sophisticated three-dimensional simulator with more realistic physics driving models will also be explored in the future.

## VI. Acknowledgment

This work was authored by the National Renewable Energy Laboratory, operated by Alliance for Sustainable Energy, LLC, for the U.S. Department of Energy (DOE) under Contract No. DE-AC36-08GO28308. Funding provided by the U.S. Department of Energy Vehicle Technologies Office under the Energy Efficient Mobility Systems program. The views expressed in the article do not necessarily represent the views of the DOE or the U.S. Government. The U.S. Government retains and the publisher, by accepting the article for publication, acknowledges that the U.S. Government retains a nonexclusive, paid-up, irrevocable, worldwide license to publish or reproduce the published form of this work, or allow others to do so, for U.S. Government purposes.

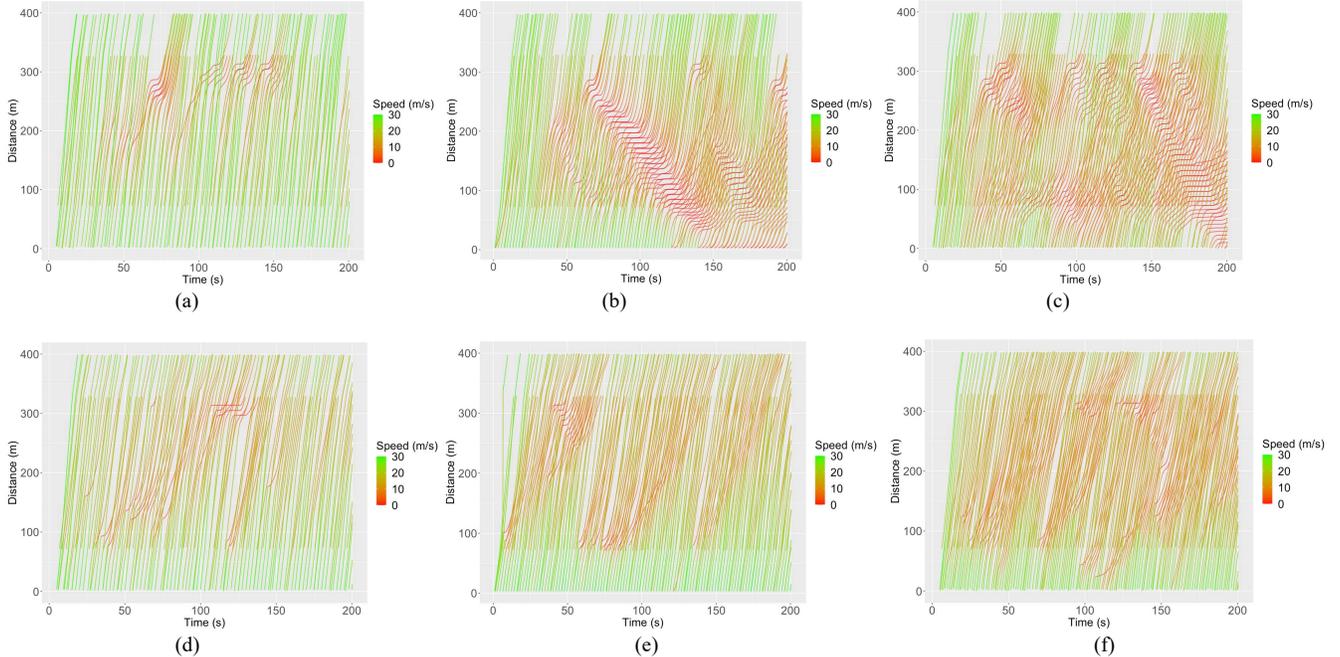

**Fig. 6.** Vehicle trajectory diagram for (a) human drivers with inflow of 900 vphpl, (b) human drivers with inflow of 1,200 vphpl, (c) human drivers with inflow of 1,500 vphpl, (d) cooperative lane changing with inflow of 900 vphpl, (e) cooperative lane changing with inflow of 1,200 vphpl, and (f) cooperative lane changing with inflow of 1,500 vphpl.

This research was performed using computational resources sponsored by the U.S. Department of Energy's Office of Energy Efficiency and Renewable Energy and located at the National Renewable Energy Laboratory.

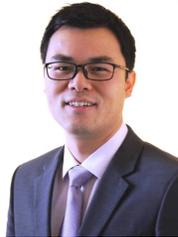

**Yi Hou** is a senior transportation researcher at the National Renewable Energy Laboratory's Center for Integrated Mobility Sciences (CIMS). He received M.S. and Ph.D. degrees in civil engineering from the University of Missouri in 2011 and 2014, respectively. The primary focus of his research revolves around machine learning, artificial intelligence (AI), and big data techniques, as well as their applications to future mobility and smart city solutions.

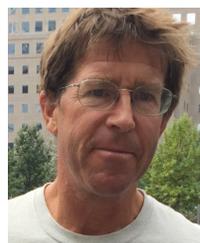

**Peter Graf** is a senior scientist in the Computational Science Center at the National Renewable Energy Laboratory. He received a B.S. in symbolic systems from Stanford University in 1989 and a Ph.D. in mathematics from the University of California at Berkeley in 2003. Dr. Graf's research brings state-of-the-art applied math, computing, and AI to bear on a wide variety of renewable energy applications. His current work focuses on reinforcement learning and quantum computing for energy systems optimization and control.